# Intermittency of inter-scale kinetic energy transfer and of energy exchange between internal and kinetic energy in turbulent premixed flames

Vladimir A. Sabelnikov[a], Andrei N. Lipatnikov[b,1],

[a]*ONERA - The French Aerospace Lab., F-91761 Palaiseau, France*
[b]*Department of Mechanics and Maritime Sciences, Chalmers University of Technology, Göteborg, 412 96, Sweden*

**Abstract**

Inter-scale kinetic energy transfer in turbulent flows is accompanied by very intense and intermittent spatial-temporal fluctuations. Such intermittency is expected to be particularly prominent in premixed flames, where heat release, density variations, dilatation, and chemical reactions are localized to spatial scales that are substantially smaller than scales of large turbulent eddies but are often larger than or comparable with Kolmogorov length scale. Nevertheless, intermittency of inter-scale energy transfer and of heat exchange between internal and kinetic energy in premixed flames has not yet been given due attention. The present work aims at bridging this knowledge gap. For this purpose, three-dimensional direct numerical simulation data obtained earlier from a statistically stationary, planar, one-dimensional lean hydrogen-air flame propagating in moderately intense turbulence are filtered using cubes of various widths. Subsequently, Probability Density Functions (PDFs) of various instantaneous flow characteristics are sampled by processing the filtered fields of velocity, pressure, density, and their spatial gradients. The sampled PDFs exhibit long tails, are highly skewed, and are characterized by a large kurtosis, thus, evidencing significant intermittency of inter-scale energy transfer and heat exchange between internal and kinetic energy in the flame.

*Keywords:* Premixed turbulent combustion; Thermal expansion; Turbulence; Intermittency; Inter-scale transfer

## I. INTRODUCTION

According to the classical statistical theory of locally isotropic and homogeneous turbulence in incompressible flows,[1-3] turbulent kinetic energy is transferred on average from large scales, where it is generated, to small scales via the turbulence cascade,[4] with the energy being dissipated due to molecular viscosity at the smallest scales. However, the modern picture of inter-scale turbulence energy transfer substantially expands this classical paradigm of energy cascade by going beyond the statistically average framework. Specifically, the local behavior of turbulence kinetic energy transfer is known to be highly intermittent in space and time, i.e., there are significant spatial-temporal fluctuations of the inter-scale energy transfer rate (both positive and negative).[5-18] Accordingly, there are localized regions with inverse energy cascade, i.e., from small scales to large scales. This phenomenon is often referred to as backscatter. The net average cascade is a result of downscale/direct and upscale/backscatter energy transfer. Even if the classical forward cascade statistically overwhelms backscatter in many turbulent flows,[6,19-21] analysis of inter-scale kinetic energy transfer was in the focus of many studies reviewed elsewhere.[22] Their results indicate that inter-scale kinetic energy transfer is sufficiently well understood in the classical case of incompressible, homogeneous, isotropic turbulence. In compressible, homogeneous, non-reacting turbulence, due dilatational velocity fluctuations, the situation is more intricate. For instance, energy exchange between kinetic and internal energies can cause complex nonlinear interactions of vortices and, e.g., acoustic or shock waves.[23-26] In compressible reacting turbulence, the physics of inter-scale energy transfer becomes even richer.[27,28] It should be stressed that the mean velocity divergence vanishes both in statistically homogeneous non-reacting and reacting compressible turbulence.

---

[1]Corresponding author, lipatn@chalmers.se

From this perspective, inter-scale energy transfer in premixed flames, where even the mean flow is highly inhomogeneous, appears to enrich the physics of turbulence cascades even farther for the following reasons. As reviewed elsewhere,[29-31] in a typical premixed turbulent flame, heat release, density variations, dilatation, and chemical reactions are localized to spatial scales comparable with laminar flame thickness, which is well below 1 mm under atmospheric conditions. These scales are substantially smaller than scales of large turbulent eddies but are often larger than or comparable with Kolmogorov length scale.[2,3] Therefore, the energy exchange between the internal and kinetic energies in premixed flames is associated with small scales, contrary to injection of kinetic energy at large scales in incompressible turbulence. Moreover, combustion occurs in highly compressible flows, with the mean dilatation being (i) not only sufficiently large when compared to velocity gradients in the incoming turbulence, (ii) but also positive, contrary to a typical non-reacting compressible flow. The latter peculiarity (positive and large mean dilatation rate) of premixed turbulent flames when compared to compressible non-reacting turbulence (where dilatation rate vanishes after averaging) can have a strong impact on the intermittency of the inter-scale energy transfer. For instance, the injection of kinetic energy and the energy exchange are strongly influenced by pressure-dilatation term in transport equations for these two sorts of energy, see Eqs. (1)-(4) below, with this term being directly connected with the dilatational turbulent motions.

Accordingly, over the past decade, the phenomenon of combustion-induced backscatter was explored by several research groups by analyzing both numerical data obtained in Direct Numerical Simulation (DNS) studies of premixed turbulent flames stabilized in simple flow configurations[32-42] and experimental data obtained from swirl flames.[43,44] However, other important aspects of intermittency of inter-scale energy transfer were beyond the focus of the cited papers. For instance, the most common approaches to studying intermittency of inter-scale energy transfer[2,3,45] deal either with a Probability Density Function (PDF) of a quantity relevant to the transfer or with images of iso-surfaces of sub-filter scale (SFS) flux. However, the present authors are not aware of the applications of any of these two major diagnostic techniques to analyzing data obtained from a premixed flame.

Accordingly, the present work aims at bridging this knowledge gap by applying the former (PDF) techniques to analyzing unsteady, three-dimensional DNS data obtained recently by Dave et al.[46,47] from a statistically stationary, planar, and one-dimensional complex-chemistry lean hydrogen-air flame propagating in moderately intense turbulence in a box. These DNS data were already explored by the present authors by averaging the data over transverse planes and time.[48-53] In the present work, intermittency of inter-scale energy transfer is investigated by spatially filtering the raw DNS data, followed by sampling various PDFs and their moments by processing unsteady, three-dimensional filtered fields of velocity, density, pressure, and their spatial gradients.

In the next section, the DNS attributes and applied numerical diagnostic techniques are summarized. Numerical results are reported and discussed in Sec. III, followed by conclusions.

## II. DNS ATTRIBUTES AND DATA ANALYSES

### A. DNS data

The DNS data were obtained[46,47] from an unconfined, statistically one-dimensional and planar, lean (the equivalence ratio $\phi = 0.81$) and slightly preheated (unburned gas temperature $T_u = 310$ K) H$_2$-air flame propagating in a box (19.18 × 4.8 × 4.8 mm) meshed using a uniform grid of 960 × 240 × 240 cells. The simulations were performed adopting an open-access PENCIL code,[54] the mixture-averaged transport model implemented into it, and a detailed chemical mechanism (21 reactions, 9 species) by Li et al.[55] The laminar flame speed $S_L = 1.84$ m/s, thickness $\delta_L = (T_b - T_u)/\max\{|\nabla T|\} = 0.36$ mm, and time scale $\tau_f = \delta_L/S_L = 0.20$ ms, where subscripts $u$ and $b$ designate unburned and burned mixture, respectively.

Homogeneous isotropic turbulence was pre-generated using forcing at low wavenumbers in a separate cube with the fully periodic boundary conditions.[46] The generation process was performed until a statistically stationary stage was reached. The obtained turbulence displays the Kolmogorov-Obukhov

5/3 spectrum[2,3] and is characterized by the rms velocity $u' = 6.7$ m/s, integral length scale $L = 3.1$ mm, turbulent Reynolds number $Re_t = u'L/\nu_u = 950$, Kolmogorov length scale $\eta_K = (\nu_u^3/\langle\varepsilon\rangle)^{1/4} = 0.018$ mm, integral and Kolmogorov time scales $\tau_t = L/u' = 0.46$ ms and $\tau_K = (\nu_u/\langle\varepsilon\rangle)^{1/2} = 0.015$ ms, respectively. Here, $\langle\varepsilon\rangle = 2\nu_u S_{ij} S_{ij}$ designates the rate of dissipation of turbulent kinetic energy, averaged over the cube; $\nu$ is kinematic viscosity; $S_{ij} = 0.5(\partial u_i/\partial x_j + \partial u_j/\partial x_i)$ is the rate-of-strain tensor; and the Einstein summation convention applies to repeated indexes. Accordingly, the Damköhler number $Da = \tau_t/\tau_f = 2.35$ and the number $(\delta_L/\eta_K)^2$, which is sometimes associated with Karlovitz number, is as large as 400. Note, that a more appropriate Karlovitz number $Ka = \tau_f/\tau_K = 13$ is significantly less, because $S_L \delta_L \gg \nu_u$ in lean H$_2$-air flames[56] but is still significantly larger than unity.

When running combustion simulations, the pre-generated turbulence was injected into the computational domain through the left boundary and decayed along the mean flow direction $x$ (symmetry boundary conditions were set at transverse boundaries). Accordingly, $u' = 3.3$ m/s and $Ka = 3.3$ at the leading edge of the mean flame brush, associated with the transverse-averaged value of the fuel-based combustion progress variable $c_F$ equal to 0.01. Nevertheless, the turbulence length scales evaluated at the inlet boundary and at the leading edge are roughly equal, i.e., $(\delta_L/\eta_K)^2$ is still about 400 at the leading edge. Here, $c_F = (Y_F - Y_{F,u})/(Y_{F,b} - Y_{F,u})$ is defined using the fuel mass fraction $Y_F$ to satisfy a constraint of $0 \leq c_F \leq 1$, whereas local values of temperature-based combustion progress variable can be larger than unity due to differences in molecular diffusivities of heat, H$_2$, and O$_2$.[57,58]

**B. Mathematical background**

Specifically, the present study aims at analyzing (i) terms that describe inter-scale energy transfer and (ii) pressure-dilatation terms in transport equations for filtered kinetic energy $\tilde{k}(\mathbf{x},t) = \widetilde{u_k u_k}/2$, resolved kinetic energy $\tilde{k}_{res}(\mathbf{x},t) = \tilde{u}_k \tilde{u}_k/2$, subgrid-scale (sgs) energy $\tilde{k}_{sgs}(\mathbf{x},t) = \tilde{k} - \tilde{k}_{res}$, and filtered total (sensible and chemical) internal energy $\tilde{e}(\mathbf{x},t)$. Such equations are widely used in Large Eddy Simulation (LES) research into turbulent flows and are well known to read[59-62]

$$\frac{\partial}{\partial t}(\bar{\rho}\tilde{k}) + \frac{\partial}{\partial x_k}(\bar{\rho}\tilde{u}_k\tilde{k}) + \frac{\partial J_k}{\partial x_k} = \overline{p\frac{\partial u_k}{\partial x_k}} - \bar{\rho}\tilde{\varepsilon}, \tag{1}$$

$$\frac{\partial}{\partial t}(\bar{\rho}\tilde{k}_{res}) + \frac{\partial}{\partial x_k}(\bar{\rho}\tilde{u}_k\tilde{k}_{res}) + \frac{\partial J_{res,k}}{\partial x_k} = -\Pi - \Lambda + \bar{p}\frac{\partial \bar{u}_k}{\partial x_k} + (\tilde{u}_k - \bar{u}_k)\frac{\partial \bar{\tau}_{v,jk}}{\partial x_j} - \bar{\tau}_{v,jk}\bar{S}_{jk}, \tag{2}$$

$$\frac{\partial}{\partial t}(\bar{\rho}\tilde{k}_{sgs}) + \frac{\partial}{\partial x_k}(\bar{\rho}\tilde{u}_k\tilde{k}_{sgs}) + \frac{\partial J_{sgs,k}}{\partial x_k}$$
$$= \Pi + \Lambda + \overline{p\frac{\partial u_k}{\partial x_k}} - \bar{p}\frac{\partial \bar{u}_k}{\partial x_k} - (\tilde{u}_k - \bar{u}_k)\frac{\partial \bar{\tau}_{v,jk}}{\partial x_j} - \overline{\tau_{v,jk}S_{jk}} + \bar{\tau}_{v,jk}\bar{S}_{jk}, \tag{3}$$

$$\frac{\partial}{\partial t}(\bar{\rho}\tilde{e}) + \frac{\partial}{\partial x_k}(\bar{\rho}\tilde{u}_k\tilde{e}) + \frac{\partial J_{e,k}}{\partial x_k} = -\overline{p\frac{\partial u_k}{\partial x_k}} + \bar{\rho}\tilde{\varepsilon}, \tag{4}$$

respectively. Here, $t$ is time; $x_k$ and $u_k$ are cartesian coordinates and components of flow velocity vector, respectively; $\rho$ and $p$ are density and pressure, respectively; overline and overtilde refer to filtered and Favre-filtered (density-weighted) quantities, respectively, i.e., $\tilde{q} \equiv \overline{\rho q}/\bar{\rho}$. For brevity, the viscous stress tensor $\tau_{v,jk}$ and the fluxes $J_k, J_{res,k}, J_{sgs,k}$, and $J_{e,k}$ are not specified here, because they are beyond the focus of the present study. The reader interested in expressions for this tensor and these fluxes is referred to Refs. [59-62].

The focus of the present analysis is placed on:

(i) instantaneous rate $\Pi \equiv -\bar{\rho}(\widetilde{u_iu_j} - \tilde{u}_i\tilde{u}_j)(\partial\tilde{u}_i/\partial x_j)$ of inertial transfer of kinetic energy between resolved and subgrid scales, see Eqs. (2) and (3), which involve this term with opposite signs,

(ii) instantaneous baropycnal work[61-63] $\Lambda \equiv (\tilde{\mathbf{u}} - \bar{\mathbf{u}}) \cdot \nabla\bar{p}$, which also appears in Eqs. (2) and (3) with opposite signs,

(iii) instantaneous resolved (large-scale) pressure-dilatation term $\bar{p}\overline{\nabla \cdot \mathbf{u}}$, see Eq. (2),

(iv) instantaneous pressure-dilatation correlation $\overline{p\nabla \cdot \mathbf{u}}$, see Eqs. (1) and (4), and

(v) instantaneous unresolved (small-scale) pressure-dilatation term $\overline{p\nabla \cdot \mathbf{u}} - \bar{p}\nabla \cdot \bar{\mathbf{u}}$, see Eq. (3).

Henceforth, large-scale and resolved quantities are considered to be equivalent, as well as unresolved and small-scale quantities.

In a general case, the local pressure-dilatation term $p\nabla \cdot \mathbf{u}$ can be either positive or negative and results in a bidirectional (reversible) exchange between internal and kinetic energies, whereas kinetic energy dissipation rate $\rho\varepsilon$ leads to one-way transfer from kinetic to internal energy. In Eqs. (1) and (4), the filtered term $\overline{p\nabla\cdot\mathbf{u}}$ describes the energy transfer between Favre-filtered internal energy $\tilde{e}$ and kinetic energy $\tilde{k}$. In Eq. (2), the large-scale pressure-dilatation term $\bar{p}\nabla\cdot\bar{\mathbf{u}}$ is responsible for energy exchange between filtered internal energy and kinetic energy $k_{res}$ of filtered (resolved) velocity field. In Eq. (3), the small-scale pressure-dilatation term $\overline{p\nabla\cdot\mathbf{u}} - \bar{p}\nabla\cdot\bar{\mathbf{u}}$ is associated with energy exchange between filtered internal energy and the subgrid scale kinetic energy $\tilde{k}_{sgs}$.

The terms $\Pi$ and $\Lambda$ in Eqs. (2) and (3) are Galilean invariants, as discussed in detail elsewhere,[64,65] and describe energy transfer across scales. These two terms require modeling. On the contrary, the term $\bar{p}\nabla\cdot\bar{\mathbf{u}}$ (i) couples filtered internal energy $\tilde{e}$ and large-scale kinetic energy $\tilde{k}_{res}$, (ii) is closed, (iii) is controlled by large-scale fields, and, therefore, (iv) cannot transfer energy over scales. The small-scale pressure-dilatation term $\overline{p\nabla\cdot\mathbf{u}} - \bar{p}\nabla\cdot\bar{\mathbf{u}}$ cannot transfer energy across scales either.[61,62] In the case of a constant density, both $\Lambda$ and $\nabla\cdot\mathbf{u}$ vanish, and inter-scale energy transfer is solely controlled by $\Pi$, with $\Pi > 0$ in the case of the classical Richardson-Kolmogorov cascade.

**C. Diagnostic techniques**

Raw data stored in the DNS database[46,47] were filtered over a cube (a top-hat filter), with three filter widths being probed: $\Delta = 0.22\delta_L$, $\Delta = 0.44\delta_L$, and $\Delta = 0.88\delta_L$. Subsequently, for various quantities $q$, obtained filtered fields $\bar{q}(\mathbf{x},t)$ or $\tilde{q}(\mathbf{x},t)$ were further analyzed by building three-dimensional PDFs $P(\bar{q},\xi,\zeta)$ so that (i) $|\bar{q}(\mathbf{x},t) - \psi| \leq d\psi$ or $|\tilde{q}(\mathbf{x},t) - \psi| \leq d\psi$, respectively, with 100 bins being used for the sampling variable $\psi$, (ii) $|\bar{c}_F(\mathbf{x},t) - \xi| \leq 0.05$, and (iii) $|\langle c_F\rangle(x,t) - \zeta| \leq 0.05$, with nine bins being used for both sampling variable $\xi$ associated with the filtered combustion progress variable $0 \leq \bar{c}_F(\mathbf{x},t) \leq 1$ and sampling variable $\zeta$ associated with the transverse-averaged combustion progress variable $\langle c_F\rangle(x,t)$, e.g., $\zeta_1, \zeta_2, \ldots, \zeta_9$ were equal to 0.1, 0.2, …, 0.9, respectively. Here, $x$-axis is normal to the mean flame surface. Such PDFs were sampled from 56 snapshots stored each 5 μs over $2.8 \leq t/\tau_t = \leq 3.4$ and were averaged over time. During this time interval, statistical stationarity of the flame propagation is reached, e.g., turbulent burning velocity oscillates weakly around a steady value.[48]

Reported in the present paper are the following shifted PDFs:

$$\mathcal{P}(s) = \sigma b^{-1} \int_{\zeta_1}^{\zeta_2} \int_{\xi_1}^{\xi_2} P(q,\xi,\zeta) d\xi \, d\zeta, \tag{5}$$

$$s = \frac{q-\mu}{\sigma}, \tag{6}$$

$$b = \int_{\zeta_1}^{\zeta_2} \int_{\xi_1}^{\xi_2} \int_{\psi_1}^{\psi_2} P(\psi,\xi,\zeta) d\psi \, d\xi \, d\zeta, \tag{7}$$

$$\mu = b^{-1} \int_{\zeta_1}^{\zeta_2} \int_{\xi_1}^{\xi_2} \int_{\psi_1}^{\psi_2} \psi P(\psi,\xi,\zeta) d\psi \, d\xi \, d\zeta, \tag{8}$$

$$\sigma^2 = b^{-1} \int_{\zeta_1}^{\zeta_2} \int_{\xi_1}^{\xi_2} \int_{\psi_1}^{\psi_2} (\psi-\mu)^n P(\psi,\xi,\zeta) d\psi \, d\xi \, d\zeta, \tag{9}$$

where $\psi_1$ and $\psi_2$ are set so that the PDF vanishes at $q < \psi_1$ or $q > \psi_2$ and $\zeta_1 = 0.05$ and $\zeta_2 = 0.95$, i.e., the PDFs were sampled from the entire flame brush volume, where the filtered combustion progress variable satisfied the following constraint $0.05 \leq \bar{c}_F(\mathbf{x},t) \leq 0.95$. As far as $\xi_1$ and $\xi_2$ are concerned, they were set equal either to 0.05 and 0.95, respectively, or to $\xi^* - 0.05$ and $\xi^* + 0.05$, respectively, with $\xi^*$ being reported in figure legends and/or captions.

The sampled PDFs are also quantified using their skewness $\mu_3/\sigma^3$ and kurtosis $\mu_4/\sigma^4$, evaluated as follows:

$$\mu_n = b^{-1} \int_{\zeta_1}^{\zeta_2} \int_{\xi_1}^{\xi_2} \int_{\psi_1}^{\psi_2} (q-\mu)^n P(\psi,\xi,\zeta) d\psi \, d\xi \, d\zeta, \tag{10}$$

where $n = 3$ or $4$, respectively.

Besides the PDFs, the following time-averaged, doubly conditioned first moments $\langle q | \bar{c}_F(\mathbf{x},t) = \xi | \langle c_F \rangle (x,t) = \zeta \rangle (\xi,\zeta)$, were directly sampled from the filtered fields $\bar{q}(\mathbf{x},t)$ or $\tilde{q}(\mathbf{x},t)$ by adopting constraints of $|\bar{c}_F(\mathbf{x},t) - \xi| \leq 0.05$ and $|\langle c_F \rangle (x,t) - \zeta| \leq 0.05$.

## III. RESULTS AND DISCUSSION

Figure 1 shows spatial variations of normalized (using $\rho_u \delta_L / S_L^3$, where $S_L$ and $\delta_L$ are laminar flame speed and thickness, respectively; $\rho_u$ is unburned gas density) time- and transverse-averaged values $\langle \cdot \rangle$ of $\Pi$ (black solid lines), $\Lambda$ (blue dashed lines), and $\overline{p\nabla \cdot \mathbf{u}} - \bar{p}\overline{\nabla \cdot \mathbf{u}}$ (red dotted-dashed lines) along the normal to the mean flame brush, with spatial dependencies of these mean values being transformed to $\langle c_F \rangle$-dependecies by taking an advantage of a monotonous increase in the mean combustion progress variable $\langle c_F \rangle(x)$ with the axial distance $x$. Since results obtained using the smallest and largest filters are similar, cf. Figs. 1a and 1b, results computed with medium $\Delta = 0.44 \delta_L$ are not reported for brevity. A decrease in the magnitude of the mean quantities with decreasing filter width is associated with the fact that the magnitudes of $\widetilde{u_i u_j} - \tilde{u}_i \tilde{u}_j$, $\tilde{\mathbf{u}} - \bar{\mathbf{u}}$, and $\overline{p\nabla \cdot \mathbf{u}} - \bar{p}\overline{\nabla \cdot \mathbf{u}}$ tend to zero as $\Delta \to 0$ (if a cubic filter degenerates to a point, each of these three terms vanishes).

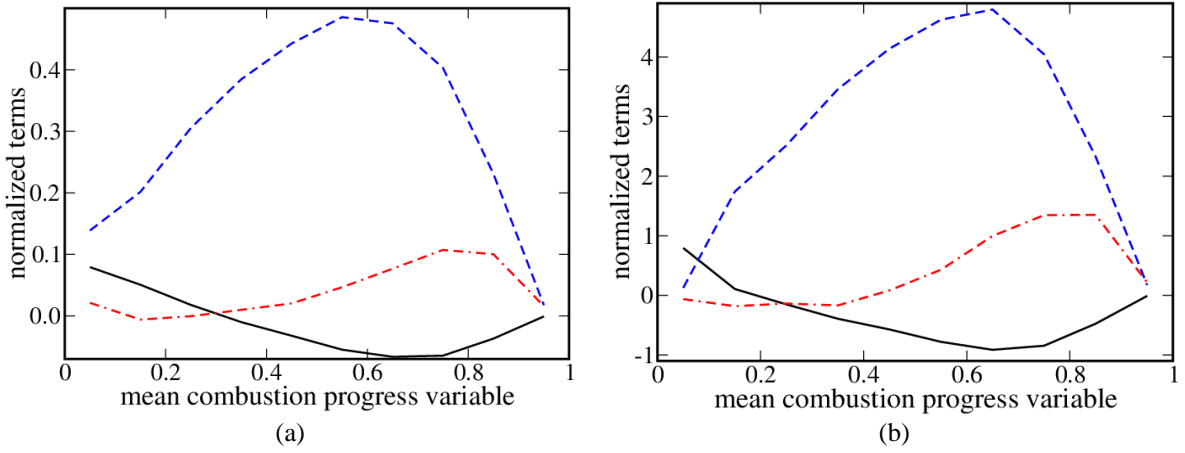

**Fig. 1**. Spatial variations of time- and transverse-averaged values of filtered fields of $\Pi$ (black solid lines), $\Lambda$ (blue dashed lines), and $\overline{p\nabla \cdot \mathbf{u}} - \bar{p}\overline{\nabla \cdot \mathbf{u}}$ (red dotted-dashed lines) along the normal to mean flame brush. All these quantities are normalized using $\rho_u \delta_L / S_L^3$. (a) $\Delta = 0.22 \delta_L$, (b) $\Delta = 0.88 \delta_L$.

The following trends shown in Fig. 1 are worth noting. First, magnitudes of $\langle \Pi \rangle$ and $\langle \overline{p\nabla \cdot \mathbf{u}} - \bar{p}\overline{\nabla \cdot \mathbf{u}} \rangle$ are comparable, whereas their signs are often opposite. Specifically, $\langle \Pi \rangle$ is positive at $\langle c \rangle < 0.3$, thus, indicating direct cascade in this leading zone of mean flame brush. However, due to the influence of combustion-induced thermal expansion $\langle \Pi \rangle < 0$ at larger $\langle c \rangle$, thus, indicating backscatter. Such a backscatter was already reported in earlier DNS studies of premixed turbulent flames.[32-41] The small-scale pressure-dilatation term $\langle \overline{p\nabla \cdot \mathbf{u}} - \bar{p}\overline{\nabla \cdot \mathbf{u}} \rangle$ is positive at $\langle c \rangle > 0.3$, indicating transfer of internal energy to subgrid scale motions, but can be negative at lower $\langle c \rangle$, indicating energy transfer in the opposite direction.

Second, $\langle \Lambda \rangle$ is always positive, with its magnitude being significantly higher than magnitudes of two other mean terms, at least, if $\langle c \rangle > 0.1$. Accordingly, $\langle \Pi \rangle + \langle \Lambda \rangle$ is always positive, i.e., these two components of SFS transfer, considered jointly, yield direct cascade.

Figure 2 shows spatial variations of normalized (using $\rho_u \delta_L/S_L^3$) mean, i.e., $\langle \bar{p}\overline{\nabla \cdot \mathbf{u}} \rangle$, see black solid line, and conditioned, i.e., $\langle \bar{p}\overline{\nabla \cdot \mathbf{u}}|\xi \rangle$, see color broken lines, values of the filtered large-scale pressure-dilatation term $\bar{p}\overline{\nabla \cdot \mathbf{u}}$. Numbers near color curves report values $\xi$ of the filtered combustion progress variable $\bar{c}_F(\mathbf{x}, t)$ that the term is conditioned to. Results obtained using the two larger $\Delta$ are similar and are not reported for brevity. However, it is worth noting that, contrary to the terms presented in Fig. 1, the magnitude of the mean large-scale pressure-dilatation $\langle \bar{p}\overline{\nabla \cdot \mathbf{u}} \rangle$ is weakly affected by $\Delta$, because this term does not involve differences between local and filtered quantities, which (differences) vanish as $\Delta \to 0$. It is also worth noting that dependencies of $\langle \bar{p}\overline{\nabla \cdot \mathbf{u}} \rangle$ and $\langle \bar{p}\nabla \cdot \bar{\mathbf{u}} \rangle$ on $\langle c_F \rangle$ are indistinguishable (not shown), as expected.

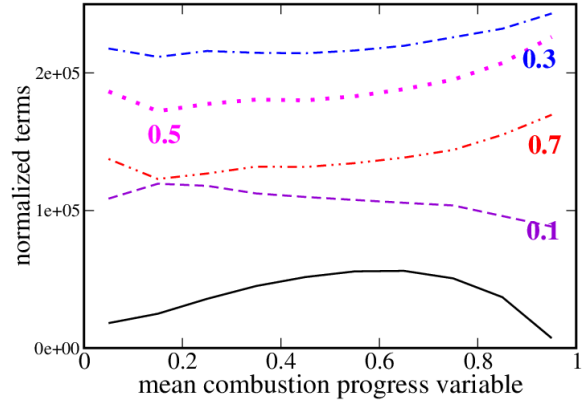

**Fig. 2**. Spatial variations of mean (black solid line) and conditioned (color broken lines) values of normalized filtered large-scale pressure-dilatation field $\bar{p}\overline{\nabla \cdot \mathbf{u}}(\rho_u \delta_L/S_L^3)$ along the normal to mean flame brush. Numbers near color curves show values $\xi$ of the filtered combustion progress variable $\bar{c}_F(\mathbf{x}, t)$ that the term $\langle \bar{p}\overline{\nabla \cdot \mathbf{u}}|\xi \rangle$ is conditioned to. $\Delta = 0.22\delta_L$.

Figure 2 shows that the mean large-scale pressure-dilatation term $\langle \bar{p}\overline{\nabla \cdot \mathbf{u}} \rangle$ and its conditioned counterpart $\langle \bar{p}\overline{\nabla \cdot \mathbf{u}}|\xi \rangle$ are always positive indicating transfer of internal energy to kinetic energy of resolved large-scale motions. Moreover, the magnitude of this term is much larger than the magnitude of the three other terms addressed in Fig. 1. The point is that, at a low Mach number typical for a free turbulent premixed flame, pressure variations within the flame brush are small, i.e., $|p(\mathbf{x}, t) - p_0| \ll p_0$. Therefore, the magnitude of $\langle \bar{p}\overline{\nabla \cdot \mathbf{u}} \rangle$ is on the order of $p_0 \langle \overline{\nabla \cdot \mathbf{u}} \rangle$ and this term is large, because it is proportional to the characteristic pressure $p_0$. For the same reasons, $\langle\overline{p\nabla\cdot\mathbf{u}}\rangle \approx \langle\bar{p}\overline{\nabla\cdot\mathbf{u}}\rangle \approx p_0\langle\nabla\cdot\mathbf{u}\rangle$ (not shown for brevity) and $|\langle\overline{p\nabla\cdot\mathbf{u}} - \bar{p}\overline{\nabla\cdot\mathbf{u}}\rangle| \ll \langle\bar{p}\overline{\nabla\cdot\mathbf{u}}\rangle$, cf. curves shown in red dotted-dashed line in Fig. 1a and in black solid line in Fig. 2. Accordingly, henceforth, any feature of $\overline{p\nabla\cdot\mathbf{u}}$ is considered to be a feature of $\bar{p}\overline{\nabla\cdot\mathbf{u}}$ also and vice versa. Nevertheless, the small-scale pressure-dilatation term $\langle\overline{p\nabla\cdot\mathbf{u}} - \bar{p}\overline{\nabla\cdot\mathbf{u}}\rangle$ is finite, see curve plotted in red dotted-dashed line in Fig. 1a, and should not be disregarded, because it plays a role in the transport equation for unresolved kinetic energy, see Eq. (3).

Furthermore, comparison of curves plotted in black solid and blue dashed lines in Fig. 1a with curve plotted in black solid line in Fig. 2, shows that the mean large-scale pressure-dilatation term $\langle\bar{p}\overline{\nabla\cdot\mathbf{u}}\rangle$ dominates the r.h.s. of Eq. (2), i.e., energy flux to resolved motions is mainly controlled by transfer of resolved internal energy, whereas inter-scale energy transfer between resolved and subgrid scale motions, i.e., terms $\Pi$ and $\Lambda$ on the r.h.s. of Eq. (2) or (3), affects weakly (substantially) the resolved (subgrid scale) kinetic energy.

Finally, Fig. 2 also shows that magnitudes of the conditioned terms $\langle\bar{p}\overline{\nabla\cdot\mathbf{u}}|\xi\rangle$ can be significantly larger than magnitude of the mean term $\langle\bar{p}\overline{\nabla\cdot\mathbf{u}}\rangle$, with this difference being most prominent at $\bar{c}_F(\mathbf{x},t) = 0.3$ and 0.5, see curves plotted in blue dotted-double-dashed and magenta dotted lines, respectively. The point is that, despite a sufficiently high Karlovitz number and a small Kolmogorov length scale $\eta_K \ll \delta_L$, the studied flame statistically retains the local structure of the counterpart unperturbed laminar premixed flame, as demonstrated in recent papers.[48,49,51,52] Accordingly, the local dilatation $\nabla\cdot\mathbf{u}$ peaks close to $c_F = 0.3$ both in that laminar flame and in the studied turbulent flame.[52] Therefore, $\langle\bar{p}\overline{\nabla\cdot\mathbf{u}}|\xi\rangle$ is the largest at $\xi = 0.3$ in Fig. 2, see curve plotted in blue dotted-double-dashed line.

Figure 3 shows that variations in the conditioned values $\langle\cdot|\xi\rangle$ with the sampling variable $\xi$ can be even much stronger for certain other considered terms. Indeed, comparison of curves plotted in cyan dotted-double-dashed and red solid lines in Fig. 3 with curve plotted in blue dashed line in Fig. 1a shows that magnitudes of the conditioned terms $\langle\Lambda|\xi = 0.1\rangle$ and, especially, $\langle\Lambda|\xi = 0.5\rangle$ are substantially larger than magnitude of the mean term $\langle\Lambda\rangle$. Such a difference is much more prominent for the small-scale pressure-dilatation term $\overline{p\nabla\cdot\mathbf{u}} - \bar{p}\overline{\nabla\cdot\mathbf{u}}$, cf. curves plotted in violet dotted-dashed and orange dashed lines in Fig. 3 with curve plotted in red dotted-dashed line in Fig. 1a. As a result, magnitudes of the conditioned terms $\langle \overline{p\nabla\cdot\mathbf{u}} - \bar{p}\overline{\nabla\cdot\mathbf{u}}|\xi = 0.1\rangle$ and $\langle \overline{p\nabla\cdot\mathbf{u}} - \bar{p}\overline{\nabla\cdot\mathbf{u}}|\xi = 0.5\rangle$ are larger than magnitudes of the conditioned terms $\langle \Lambda|\xi = 0.1\rangle$ and $\langle \Lambda|\xi = 0.5\rangle$, respectively, cf. curves plotted in violet dotted-dashed and cyan dotted-double-dashed lines or curves plotted in orange dashed and red solid lines in Fig. 3. On the contrary, Fig. 1a shows that magnitude of $\langle \Lambda \rangle$ is significantly larger than magnitude of the small-scale pressure-dilatation term $\langle \overline{p\nabla\cdot\mathbf{u}} - \bar{p}\overline{\nabla\cdot\mathbf{u}} \rangle$. Such opposite inequalities between magnitudes of conditioned terms and magnitudes of mean terms are associated with the fact that dilatation is localized to thin zones in a typical premixed turbulent flame and, in particular, in the studied flame.[50] When averaging is performed over a transverse plane, probability of finding such zones is low and the mean term is small. When averaging is performed over volumes characterized by $\bar{c}_F(\mathbf{x}, t) = 0.1$ or 0.5, probability of finding high dilatation $\nabla\cdot\mathbf{u}$ is significant and both conditioned subterms $\langle \overline{p\nabla\cdot\mathbf{u}}|\xi \rangle$ and $\langle \bar{p}\overline{\nabla\cdot\mathbf{u}}|\xi \rangle$ are large when compared to their mean counterparts. Since the baropycnal work $\Lambda$ is controlled by large-scale flow characteristics, it is significantly less sensitive to averaging method when compared to the pressure-dilatation terms $\overline{p\nabla\cdot\mathbf{u}}$ and $\bar{p}\overline{\nabla\cdot\mathbf{u}}$.

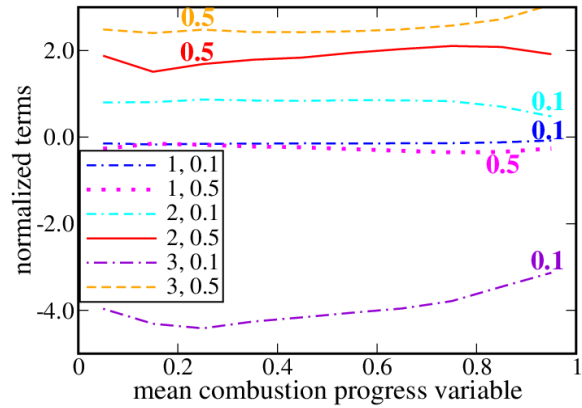

**Fig. 3**. Spatial variations of conditioned values of normalized (using $\rho_u \delta_L / S_L^3$) filtered fields along the normal to mean flame brush. The first number in each legend refers to terms (1) $\Pi$, (2) $\Lambda$, or (3) $\overline{p\nabla\cdot\mathbf{u}} - \bar{p}\overline{\nabla\cdot\mathbf{u}}$. The second numbers in each legend and numbers near curves show values $\xi$ of the filtered combustion progress variable $\bar{c}_F(\mathbf{x}, t)$ that the terms $\langle \cdot |\xi \rangle$ are conditioned to. $\Delta = 0.22 \delta_L$.

Curves plotted in violet dotted-dashed and orange dashed lines in Fig. 3 show that the signs of $\langle \overline{p\nabla\cdot\mathbf{u}} - \bar{p}\overline{\nabla\cdot\mathbf{u}}|\xi = 0.1\rangle$ and $\langle \overline{p\nabla\cdot\mathbf{u}} - \bar{p}\overline{\nabla\cdot\mathbf{u}}|\xi = 0.5\rangle$ are opposite. This difference can be explained by recalling that the studied flame statistically retains the local structure of the counterpart unperturbed laminar premixed flame.[48,49,51,52] In the latter flame, pressure monotonously decreases from unburned to burned sides, whereas dilatation grows from zero to a peak value reached at $c_F \approx 0.3$ and decreases with further increasing $c_F$. Accordingly, correlation between pressure and dilatation should be negative and positive at $c_F < c_F^*$ and $c_F > c_F^*$, respectively, with $c_F^* \approx 0.3$. These simple reasoning explain opposite signs of $\langle \overline{p \nabla \cdot \mathbf{u}} - \bar{p} \overline{\nabla \cdot \mathbf{u}} | \xi = 0.1 \rangle < 0$ and $\langle \overline{p \nabla \cdot \mathbf{u}} - \bar{p} \overline{\nabla \cdot \mathbf{u}} | \xi = 0.5 \rangle > 0$.

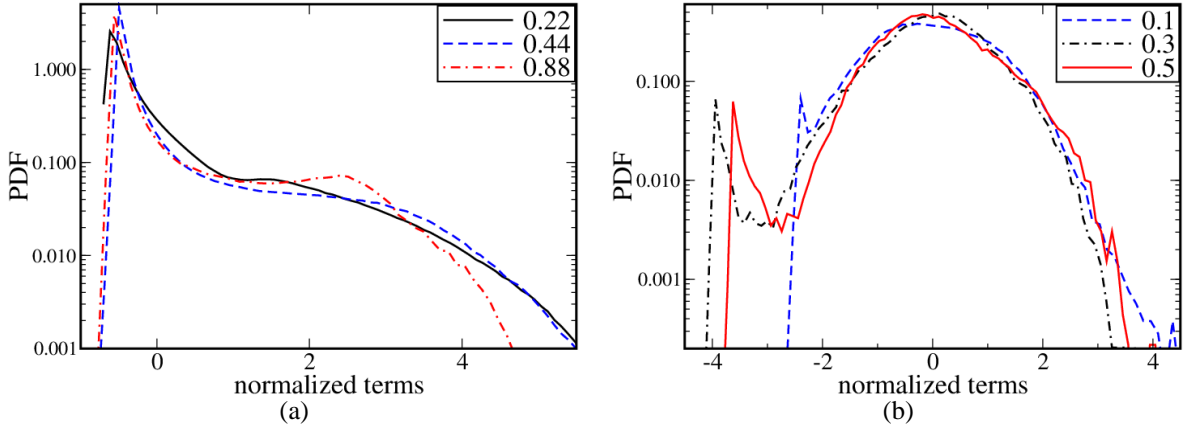

**Fig. 4**. Probability density functions $\mathcal{P}(s)$ sampled at either (a) $0.05 < \bar{c}_F(\mathbf{x}, t) < 0.95$ using different normalized filter widths $\Delta/\delta_L$, specified in legends, or (b) $|\bar{c}_F(\mathbf{x}, t) - \xi^*| < 0.05$ using $\Delta/\delta_L = 0.44$, with $\xi^*$ being specified in legends. Here, $s = (T - \mu_T)/\sigma_T$, the normalized filtered term $T(\mathbf{x}, t) = (\rho_u \delta_L/S_L^3) \bar{p} \overline{\nabla \cdot \mathbf{u}}$, $\mu_T$ and $\sigma_T$ are its mean and rms values evaluated in the considered interval of $\bar{c}_F(\mathbf{x}, t)$.

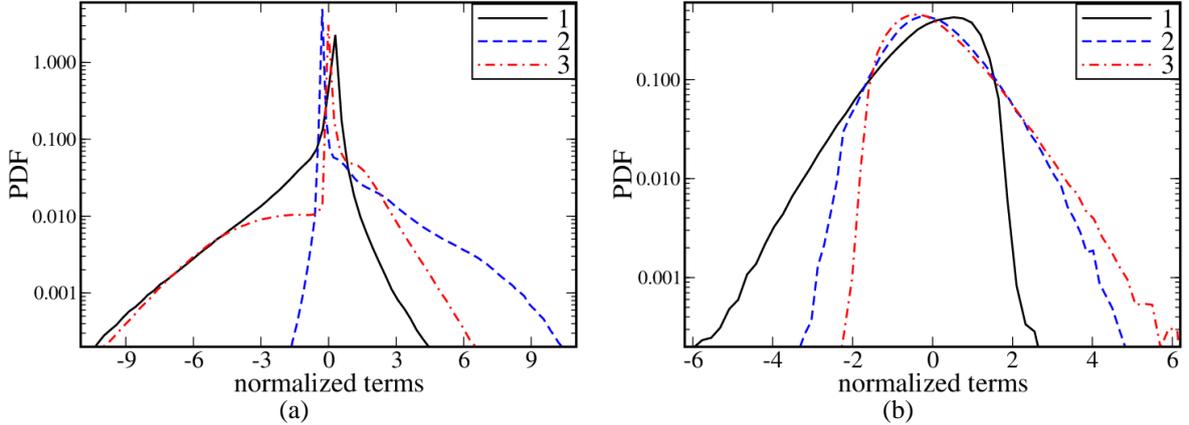

**Fig. 5**. Probability density functions $\mathcal{P}(s)$ sampled using $\Delta/\delta_L = 0.22$ at either (a) $0.05 < \bar{c}_F(\mathbf{x}, t) < 0.95$ or (b) $|\bar{c}_F(\mathbf{x}, t) - 0.5| < 0.05$. Here, $s = (T_n - \mu_{T_n})/\sigma_{T_n}$, the normalized filtered terms $T_n(\mathbf{x}, t)$ are equal to (1) $T_1 = (\rho_u \delta_L/S_L^3)\Pi$, (2) $T_2 = (\rho_u \delta_L/S_L^3)\Lambda$, and (3) $T_3 = (\rho_u \delta_L/S_L^3)(\overline{p \nabla \cdot \mathbf{u}} - \bar{p} \overline{\nabla \cdot \mathbf{u}})$, with $\mu_{T_n}$ and $\sigma_{T_n}$ being their mean and rms values evaluated in the considered interval of $\bar{c}_F(\mathbf{x}, t)$.

Since intermittency is turbulent flows is commonly emphasized by reporting PDFs of relevant quantities, let us explore intermittency of the analyzed random filtered fields by investigating their PDFs conditioned to either $0.05 < \bar{c}_F(\mathbf{x}, t) < 0.95$ or $|\bar{c}_F(\mathbf{x}, t) - \xi^*| < 0.05$ (with $\xi^* = 0.1$, 0.3, or 0.5) and sampled from the entire flame brush. Certain representative PDFs are reported in Figs. 4 and 5, with the major characteristics of all PDFs being summarized in Table I. PDFs of the filtered large-scale pressure-dilatation term term $\bar{p} \overline{\nabla \cdot \mathbf{u}}$ (or $\overline{p \nabla \cdot \mathbf{u}} \approx \bar{p} \overline{\nabla \cdot \mathbf{u}}$) are shown on Fig. 4 separately from PDFs of other terms, see Fig. 5, because this term (i) has a much higher magnitude and (ii) describe transfer of resolved internal energy to resolved and filtered motions, whereas the three other terms are associated with energy transfer to subgrid scale motions.

**TABLE I.** Standard deviation $\sigma$, skewness $\mu_3/\sigma^3$, and kurtosis $\mu_4/\sigma^4$ of PDFs for various normalized filtered terms.

| Normalized terms | $\Delta = 0.22\delta_L$ | | | $\Delta = 0.44\delta_L$ | | | $\Delta = 0.88\delta_L$ | | |
|---|---|---|---|---|---|---|---|---|---|
| | $\sigma$ | $\mu_3/\sigma^3$ | $\mu_4/\sigma^4$ | $\sigma$ | $\mu_3/\sigma^3$ | $\mu_4/\sigma^4$ | $\sigma$ | $\mu_3/\sigma^3$ | $\mu_4/\sigma^4$ |
| entire flame brush | | | | | | | | | |
| $\Pi$ | 0.12 | -4.4 | 31.5 | 0.36 | -4.2 | 26.4 | 0.97 | -3.0 | 15.8 |
| $\Lambda$ | 0.80 | 4.9 | 31.9 | 2.4 | 4.4 | 24.4 | 5.4 | 3.6 | 16.4 |
| $\overline{p\nabla\cdot\mathbf{u}} - \bar{p}\overline{\nabla\cdot\mathbf{u}}$ | 1.3 | -2.7 | 26.9 | 2.9 | -0.51 | 20.3 | 5.9 | 0.57 | 11.0 |
| $(\bar{p}\overline{\nabla\cdot\mathbf{u}}) \times 10^{-5}$ | 0.58 | 2.4 | 8.6 | 0.52 | 2.6 | 9.2 | 0.46 | 2.1 | 6.35 |
| $0.05 < \bar{c}_F(\mathbf{x},t) < 0.15$ | | | | | | | | | |
| $\Pi$ | 0.17 | -2.7 | 19.3 | 0.40 | -2.3 | 20.8 | 0.67 | -1.9 | 16.9 |
| $\Lambda$ | 0.90 | 1.2 | 4.3 | 1.8 | 1.1 | 5.7 | 2.7 | 1.8 | 14.5 |
| $\overline{p\nabla\cdot\mathbf{u}} - \bar{p}\overline{\nabla\cdot\mathbf{u}}$ | 2.6 | -0.82 | 3.6 | 4.8 | -0.31 | 3.2 | 5.6 | 0.28 | 5.0 |
| $(\bar{p}\overline{\nabla\cdot\mathbf{u}}) \times 10^{-5}$ | 0.52 | 0.39 | 2.7 | 0.38 | 0.10 | 2.8 | 0.23 | 0.08 | 3.3 |
| $0.25 < \bar{c}_F(\mathbf{x},t) < 0.35$ | | | | | | | | | |
| $\Pi$ | 0.26 | -0.86 | 4.0 | 0.71 | -0.70 | 4.6 | 1.2 | -0.74 | 4.5 |
| $\Lambda$ | 1.7 | 0.34 | 2.8 | 4.6 | -0.28 | 2.8 | 7.1 | -0.19 | 2.9 |
| $\overline{p\nabla\cdot\mathbf{u}} - \bar{p}\overline{\nabla\cdot\mathbf{u}}$ | 2.0 | -0.51 | 6.1 | 4.7 | -0.68 | 3.8 | 6.8 | 0.14 | 3.3 |
| $(\bar{p}\overline{\nabla\cdot\mathbf{u}}) \times 10^{-5}$ | 0.61 | 0.065 | 3.0 | 0.48 | -0.71 | 5.0 | 0.29 | -0.94 | 7.9 |
| $0.45 < \bar{c}_F(\mathbf{x},t) < 0.55$ | | | | | | | | | |
| $\Pi$ | 0.15 | -0.92 | 4.3 | 0.64 | -0.67 | 4.0 | 1.6 | -0.59 | 3.6 |
| $\Lambda$ | 0.83 | 0.46 | 3.6 | 3.8 | 0.21 | 3.0 | 8.6 | -0.38 | 2.9 |
| $\overline{p\nabla\cdot\mathbf{u}} - \bar{p}\overline{\nabla\cdot\mathbf{u}}$ | 1.6 | 1.0 | 4.8 | 4.6 | 0.39 | 3.7 | 6.1 | -0.14 | 3.8 |
| $(\bar{p}\overline{\nabla\cdot\mathbf{u}}) \times 10^{-5}$ | 0.52 | 0.68 | 3.8 | 0.51 | -0.21 | 4.3 | 0.33 | -0.86 | 7.8 |

Figure 4a shows that, independently of $\Delta/\delta_L$, PDFs of $\bar{p}\overline{\nabla\cdot\mathbf{u}}$ (or $\overline{p\nabla\cdot\mathbf{u}} \approx \bar{p}\overline{\nabla\cdot\mathbf{u}}$, not reported for brevity) are associated with strong intermittency. Indeed, these PDFs have heavy right tails, i.e., they are highly positively skewed, thus, implying rare but intense energy flux from internal energy to fluid motion due to positive velocity divergence generated in zones that heat release and density variations are localized to. Reverse energy flux from fluid motion to internal energy is less efficient. Low values of the PDFs along these heavy right tails indicate that the volume of the aforementioned zones is essentially smaller than the entire flame-brush volume. The PDF's skewness and kurtosis are both significant (larger than 2 and 6, respectively, see $\mu_3/\sigma^3$ and $\mu_4/\sigma^4$, respectively, in Table I), but do not show any clear trend regarding dependence of these two quantities on $\Delta/\delta_L$.

The PDFs conditioned to filtered combustion progress variable $\bar{c}_F(\mathbf{x},t)$ have significantly shorter right tails e.g., see Fig. 4b, thus, indicating that intensity of energy flux from internal energy to fluid motions is weaker within the conditioned framework. Moreover, the conditioned PDFs are characterized by a lower skewness whose magnitude is always smaller than 1.0, see Table I. Furthermore, flatness of the conditioned PDFs is also smaller when compared to the PDFs shown in Fig. 4a, but only if $\Delta/\delta_L = 0.22$ or $0.44$. If $\Delta/\delta_L = 0.88$ and $0.25 < \bar{c}_F(\mathbf{x}, t) < 0.35$ or $0.45 < \bar{c}_F(\mathbf{x}, t) < 0.55$, the conditioned PDFs of $\bar{p}\overline{\nabla \cdot \mathbf{u}}$ are characterized by a higher flatness. On the contrary, if $\Delta/\delta_L = 0.22$, the conditioned PDF flatness is sufficiently close to a value of 3, associated with the Gaussian distribution. All in all, comparison of Figs. 4a and 4b and data reported in Table I indicate that intermittency is less pronounced for the conditioned large-scale pressure-dilatation term $\bar{p}\overline{\nabla \cdot \mathbf{u}}$. This trend may be attributed to the fact that conditioned quantities are sampled from more uniform volumes.

It is of interest to note that, contrary to the PDFs presented in Fig. 4a, the conditioned PDFs plotted in Fig. 4b exhibit two peaks, with the appearance of the left peak changing the PDF skewness from positive to negative, see four cells for $\bar{p}\overline{\nabla \cdot \mathbf{u}}$, $\Delta/\delta_L = 0.44$, and (i) $0.05 < \bar{c}_F(\mathbf{x}, t) < 0.95$ ($\mu_3/\sigma^3 = 2.6 > 0$), (ii) $0.05 < \bar{c}_F(\mathbf{x}, t) < 0.15$ ($\mu_3/\sigma^3 = 0.10 > 0$), (iii) $0.25 < \bar{c}_F(\mathbf{x}, t) < 0.35$ ($\mu_3/\sigma^3 = -0.71 < 0$), or (iv) $0.45 < \bar{c}_F(\mathbf{x}, t) < 0.45$ ($\mu_3/\sigma^3 = -0.21$) in Table I. While the left peaks of the shifted PDFs $\mathcal{P}(s)$ appear at different $s = (T - \mu_T)/\sigma_T$, this difference results solely from dependence of the mean value $\mu_T$ on the intervals of $\bar{c}_F(\mathbf{x}, t)$ the PDFs are conditioned to. The unshifted conditioned PDFs $P(\bar{p}\overline{\nabla \cdot \mathbf{u}})$ always exhibit the left peaks at the lowest magnitude $|\bar{p}\overline{\nabla \cdot \mathbf{u}}|$, thus, indicating non-negligible probability of finding low filtered dilatation $|\overline{\nabla \cdot \mathbf{u}}|$ conditioned to volumes characterized by $0.05 < \bar{c}_F(\mathbf{x}, t) < 0.15$, $0.25 < \bar{c}_F(\mathbf{x}, t) < 0.35$, or $0.45 < \bar{c}_F(\mathbf{x}, t) < 0.45$. This is not surprising, because a cube with $\Delta/\delta_L = 0.44$ contains 512 DNS cells and many of them may be located in reactants or products, where dilatation almost vanishes.

Figure 5a shows that the PDFs of three other terms, i.e., $\Pi$, $\Lambda$, and $\overline{p\nabla \cdot \mathbf{u}} - \bar{p}\overline{\nabla \cdot \mathbf{u}}$, sampled at $0.05 < \bar{c}_F(\mathbf{x}, t) < 0.95$, are very different from Gaussian distribution. For all three terms, the PDFs are characterized by a very large kurtosis, which is decreased with increasing $\Delta/\delta_L$, see Table I. The PDF of $\Pi$, see curve plotted in black solid line in Fig. 5a, is characterized by a highly negative skewness and has a long upward tail, which looks like a stretched exponential tail. The PDF of small-scale pressure-dilatation term $\overline{p\nabla \cdot \mathbf{u}} - \bar{p}\overline{\nabla \cdot \mathbf{u}}$ has a long upward tail also and is negatively skewed, see curve plotted in red dotted-dashed line in Fig. 5a, but the skewness magnitude is smaller and even becomes small at $\Delta/\delta_L = 0.88$, see Table I. On the contrary, the PDF of $\Lambda$ is characterized by a highly positive skewness at all three $\Delta/\delta_L$ with the PDF tails appearing to be a combination of exponential and downward branches. All in all, the three PDFs sampled at $0.05 < \bar{c}_F(\mathbf{x}, t) < 0.95$ and shown in Fig. 5a, as well as their characteristics reported in Table I, indicate strong intermittency of the studied filtered fields.

The intermittency is less pronounced for PDFs conditioned to various $\bar{c}_F(\mathbf{x}, t)$. For instance, PDFs plotted in Fig. 5b have shorter tails when compared to the PDFs presented in Fig. 5a. Moreover, the conditioned PDFs are characterized by sufficiently low skewness and kurtosis, see Table I, with the exception of (i) the PDF of $\Pi$, conditioned to $0.05 < \bar{c}_F(\mathbf{x}, t) < 0.15$, (ii) the PDF of $\Lambda$, conditioned to $0.05 < \bar{c}_F(\mathbf{x}, t) < 0.15$ and obtained using the widest filter $\Delta/\delta_L = 0.88$.

## IV. CONCLUDING REMARKS

The reported DNS results indicate significant intermittency of inter-scale energy transfer and of energy exchange between internal and kinetic energy in turbulent premixed flames. This conclusion is based on analyses of Probability Density Functions of (i) two filtered inter-scale kinetic energy transfer terms, i.e., inertial inter-scale kinetic energy transfer and baropycnal work, and (ii) two terms describing the energy exchange between internal and kinetic energies, i.e., large-scale pressure-dilatation term responsible for the energy exchange between filtered internal energy and kinetic energy of resolved velocity field and small-scale pressure-dilatation term associated with the energy exchange between the filtered internal energy and subgrid scale kinetic energy. These PDFs exhibit long tails, are highly skewed, and are characterized by a large kurtosis, thus indicating the presence of relatively rare but vigorous ejection (splash) events. When analyzing filtered quantities conditioned to a narrow interval of filtered combustion progress variable, such intermittency effects are less prominent. The reported results can be used for assessment of SGS models that were already developed to allow for backscatter of turbulent energy and scalar variance, but in non-reacting constant-density flows, e.g., see a work by Schumann.[66] Moreover, the reported results can also be used to develop new SGS models that account for intermittency effects in Large Eddy Simulations of premixed turbulent combustion.

## ACKNOWLEDGEMENTS

ANL gratefully acknowledges the financial support provided by Chalmers Area of Advance Transport and by Swedish Research Council. VAS gratefully acknowledges the financial support provided by ONERA.

## AUTHOR DECLARATIONS

### Conflict of Interests

The authors declare that they have no known competing financial interests or personal relationships that could have appeared to influence the work reported in this paper.

### Author contributions

**Vladimir A. Sabelnikov:** Conceptualization (equal); Study design (equal); Analyses of the data (equal); Writing – original draft (equal); Writing – review and editing (equal). **Andrei N. Lipatnikov:** Conceptualization (equal); Study design (equal); Analyses of the data (equal); Writing – original draft (equal); Writing – review and editing (equal).

## DATA AVAILABILITY

The data that support the findings of this study are available from the corresponding author upon reasonable request.